\documentclass[12pt]{article}

\usepackage[backend=bibtex]{biblatex}
\usepackage{amsmath}
\usepackage{amstext}
\usepackage{amssymb}
\usepackage{graphicx}
\usepackage[english]{babel}
\usepackage[autostyle]{csquotes}
\usepackage{geometry}
\usepackage{bigfoot}
\usepackage{hyperref}
\usepackage{rotating}

\begin{document}

\title{How do fishery policies affect Hawaii's longline fishing industry? Calibrating a positive mathematical programming model
\thanks{%
This manuscript is published as \it{Natural Resource Modeling}. 2017;e12127.}}
\author{Jonathan R. Sweeney \thanks{%
Corresponding Author. \ Address: 2424 Maile Way; 542 Saunders Hall;
Honolulu, HI 96822. \ Tele: NA. E-mail: \texttt{%
jrsweene@hawaii.edu}.} \and Richard E. Howitt \and Hing Ling Chan \and Minling Pan \and PingSun Leung
}
\maketitle

\begin{abstract}

We present a vessel and target-specific positive mathematical programming model (PMP) for Hawaii's longline fishing fleet. Although common in agricultural economics, PMP modeling is rarely attempted in fisheries.  To demonstrate the flexibility of the PMP framework, we separate tuna and swordfish production technologies into three policy relevant fishing targets.  We find the model most accurately predicts vessel-specific annual bigeye catch in the WCPO, with an accuracy of 12\% to 35\%, and a correlation between 0.30 and 0.53. To demonstrate the model's usefulness to policy makers, we simulate the economic impact to individual vessels from increasing and decreasing the bigeye catch limit in the WCPO by 10\%. Our results suggest that such policy changes will have moderate impacts on most vessels, but large impacts on a few generating a fat tailed distribution. These results offer insights into the range of winners and losers resulting from changes in fishery policies, and therefore, which policies are more likely to gain widespread industry support. As a tool for fishery management, the calibrated PMP model offers a flexible and easy-to-use framework, capable of capturing the heterogeneous response of fishing vessels to evaluate policy changes. 

\end{abstract}

\newpage


\section{Introduction}
Understanding the economic impact of a proposed policy is crucial for ensuring policy objectives are met without being excessively burdensome on the regulated industry.  In fisheries, managers are often responsible for preventing over-fishing of common-pool fish stocks.  This involves developing policies that balance biological sustainability with economic impacts to the fishing industry.  To date, many tools available to managers measure economic impacts at the aggregate industry-level.  These tools conceal important information on differences between the impacts felt by individual firms or by types of vessels.  Sorting firms that benefit and those that are harmed can help managers understand the economic implications from the policy and which policies are expected to be equitable. 

We investigate individual vessel response to fishery policy changes using a vessel and target-specific positive mathematical programming (PMP) model.  This research is important for several reasons.  To the best of our knowledge, there have only been three previous attempts to apply PMP modeling to fisheries, although none have been published in a peer-reviewed journal.\footnote{Niels Vestagaard [1998] Policy Model for a Regulated Industry: From Command and Control to Property Rights in a Danish Multispecies Fishery, Dissertation Chapter.}\footnote{John Walden [2006] Applying Positive Math Programming to a Fisheries Problem: Formulating the Closed Area Model Structure, Social Sciences Branch, NEFSC, Wood Hole, MA, 02543, Unpublished Manuscript.}\footnote{Kathereen Bisack and Gisele Magnusson [2009] Modifications to the Harbor Porpoise Take Reduction Plan. Final Environmental Assessment, NOAA-NMFS Northeast Region.}  This provides an opportunity to formalize the PMP model structure for fisheries, which will serve as reference point in the literature and encourage further model development.  Given the panel data structure available for Hawaii's longline fishery, we are able to evaluate the performance of the fishery PMP model by comparing out-of-sample predictions to observations from reference years.  By calibrating a vessel and target-specific PMP model, this paper provides insights into the range of individual vessel responses to realistic policy changes.  Finally, this paper develops a flexible tool for fishery managers to evaluate heterogeneous policy impacts with relatively few data requirements.

Recent research suggests that fisher heterogeneity is particularly important in the Hawaii longline fleet. Fishers have differing attitudes toward risk (\cite{nguyen2013revenue}), make entry/exit decisions depending on individual fisher characteristics (\cite{pradhan2004modeling}), and choose remuneration schemes based on owner/operator status (Nguyen and Leung 2009).  The network position of individual fishers in the industry has also been shown to play an important role in determining outcomes (\cite{barnes2016brokerage}).  These studies taken together largely invalidate the common modeling assumption that the Hawaii longline fleet is homogeneous and can be modeled using a representative vessel (\cite{Kasaoka1989, Kasaoka1990}).   

Developing a model of individual vessel response to specific policy changes will, therefore, improve fleet-wide modeling accuracy.  For managers of Hawaii's longline fishery, this has added significance given the economic prominence of Hawaii's longline fishing fleet.  In 2013, the fleet landed 27,053 tons of fish and generated \$88.8 million gross revenues (\cite{WPacFIN2015}).  The fleet primarily targets swordfish and tuna in the Eastern Pacific and Western and Central Pacific regions.  It is the largest commercial fishing fleet by revenue in the state of Hawaii, with between 124 and 135 vessels operating from 2005 to 2013 (\cite{WPacFIN2015}). 

The geographic scale and environmental effects of the fishery have led managers to implement numerous regulatory policies.  The fishery is subject to gear restrictions, turtle bycatch caps, and annual catch limit restrictions.  In recent years, the fishery has been forced to close a number of times after these policy limits were reached.  In 2006 and 2011, the fishery targeting swordfish was closed because the turtle interaction limit was reached.  In 2009, 2010, and 2015, the fishery targeting bigeye tuna in the Western and Central Pacific Ocean was closed because the catch limit had been reached.  There is evidence that these closures may have had a dramatic economic impact on both producers and consumers in Hawaii (\cite{allen2006monitoring}).  

This paper examines how policies impact individual vessels by calibrating a vessel and target-specific PMP model for Hawaii's longline fishing fleet.  By calibrating at the vessel-specific level, we hope to capture the fleet's heterogeneous composition of vessels and heterogeneous response to policy changes.  We also account for two primary fishing technologies targeting bigeye tuna and swordfish, and two policy relevant management areas for bigeye tuna, one in the Eastern Pacific Ocean (EPO) and the other in the Western Central Pacific Ocean (WCPO).  In order to make our model computationally feasible, and economically tractable we make several assumptions.  First, we assume that vessels are profit maximizing.  We feel this assumption is appropriate when modeling a large commercial fishing fleet.  Second, we assume economic, environmental, and biological conditions are stable, and base year observations are representative of the important economic relationships in the fishery. Under these assumptions, we model the fishery using an objective function that maximizes individual vessel profit subject to fleet-wide annual catch constraints.  Individual model parameters are then calibrated to reproduce input and output levels from an observed base year (2012).  Using the calibrated model and observed catch data from 2009 to 2013, we then examine model accuracy using out-of-sample model predictions. To demonstrate the model's usefulness to fishery managers, we evaluate the impact of changing the catch limit policies for bigeye fishing in the WCPO.

Although the first application of PMP was more than 25 years ago (\cite{kasnakoglu1988concept}), the PMP framework was formalized by Howitt in 1995.  The idea was to blend mathematical programming constraints, which proved useful for modeling resource and policy constraints, with "positive" inferences based on observed input allocations and production levels from a particular base year.  This approach was notably different from previous "normative" mathematical programming models (\cite{day1961recursive, mccarl1982cropping}) in that it was able to exactly reproduce observed inputs and outputs without relying on numerous "flexibility" constraints, which are an additional set of constraints added by the researcher to artificially avoid corner solutions.  The general PMP framework can be specified using many structural forms of production and cost functions allowing for non-linearity and substitution between inputs, and can be easily calibrated using observations from a single year.  It is both consistent with microeconomic theory, and when applied to policy analysis, is able to generate smooth responses to policy adjustments.

These desirable modeling characteristics have made the PMP approach common in agricultural economic modeling.  Recent versions of regional agricultural models employing PMP include SWAP in California (\cite{howitt2012calibrating}), CAPRI in Europe (\cite{gocht2011eu}), and REAP in the US (\cite{Johansson2007}).  These models are used repeatedly to evaluate regional agricultural response to policy changes.  \cite{heckelei2012positive} and \cite{merel2014theory} provided comprehensive reviews of regional agricultural models currently using the PMP framework and recent developments in the PMP literature.  There has also been significant work on developing the economic foundations of PMP, emphasizing accurate estimation of supply elasticities to be used as priors (\cite{merel2010exact}), structurally consistent estimation of shadow values (\cite{heckelei2003estimation}), and improved calibration methods (\cite{garnache2015calibration}). 

By applying the most recent PMP framework developed by \cite{garnache2015calibration}, this paper builds on extensive literature modeling fleet dynamics of Hawaii's longline fishery using mathematical programming.\footnote{Curtis and Hicks (2000) investigated the impacts of fishery closure due to turtle interaction caps using a random utility model to account for spatial choice behavior of fishers.}  The first model by \cite{Inc.1986}, later modified by \cite{Kasaoka1989, Kasaoka1990}, applied a linear programming (LP) framework to optimally allocate fishing time across fishing regions and target species to maximize fleet-wide profits.  The results, however, did not accurately reproduce observed fishing behavior.  \cite{Miklius1990} evaluated the LP model and concluded that this shortcoming resulted from the omission of micro-level decision-making by vessel owners and operators.  To address this problem, \cite{pan2001decision} developed a two-level two-objective mathematical programming model which incorporated the behavior of fishers as well as fishery managers, including separate objectives of recreational and commercial fisheries.  Their approach produced more plausible optimal solutions, but it remained unclear whether the approximated profit maximizing behavior was representative.  The model also assumed that vessels within the fleet were homogenous and was, therefore, unable to capture the variation in vessel responses to changes in management. To address fleet heterogeneity in Hawaii's longline fishery, \cite{Yu2013} used an agent-based model.  While the agent-based model was able to capture some of the detailed behavior of individual fishers, there remained a fair amount of discrepancy between predicted and observed performances.  The agent-based approach to simulation also required significant model updating and refinement as well as specialized users to operate the software.

Our approach using the PMP framework is intended to be used by policy makers and managers, as well as academics.  The vessel and target-specific PMP model is able to capture fleet heterogeneity, separate fishing technologies and regional policies, and measure the distributional effects from changes to fishery policy.  It requires minimal data to calibrate, and is amenable to a wide range of resource and policy constraints including catch limits, and protected species interaction caps.  It is also able to exactly reproduce base year inputs, costs, revenues, and profits for individual vessels without relying on additional constraints.  For these reasons we feel it will be able to address previous modeling limitations.

This paper makes four important contributions to the literature.  First, the paper adapts the PMP framework developed for agriculture to a framework that can be applied to fisheries in general.  With only a handful of notable exceptions, research using PMP for fisheries policy analysis has been very limited.  Second, by calibrating a vessel and target-specific PMP model, we are able to demonstrate a technique to examine the heterogeneous nature of the fishing fleet and the heterogeneous responses to specific policy changes.  Previous literature on Hawaii's longline fishery has made significant progress to address fleet heterogeneity, but this paper provides a method that explicitly models individual vessels and fish targeting decisions, and requires less data and less effort to calibrate and conduct policy simulations than previous frameworks.  Third, it provides a rigorous out-of-sample evaluation of the accuracy of PMP model predictions.  Although PMP models have been used extensively for policy analysis, model predictions are rarely evaluated.  The panel data we have on Hawaii's longline fishery enable us to make out-of-sample predictions for catch and evaluate the model's predictive accuracy.  Finally, the calibrated PMP model of Hawaii's longline fishery provides a valuable tool for resource managers and policy analysts to evaluate the heterogeneous economic impacts of specific fishery policies and determine which policies are likely to encounter industry support or opposition.

\section{Data}
To calibrate the PMP model, evaluate its performance, and simulate policy outcomes, we used data from four sources.  We obtained data on individual vessel input costs for 2005 from the 2005 cost and earnings survey (\cite{Pan2015}), and for 2012 from the 2012 cost and earnings survey (\cite{Pan2015a}).  We obtained data on annual vessel catch from 2005-2013 using the dealer data from the State of Hawaii (\cite{Quach2015}).  We obtained data on annual hooks deployed from 2005-2013 from Federal logbook data (\cite{Monitoring2015}).  To evaluate out-of-sample prediction accuracy we adjusted all input and output prices to 2012 dollars using the Consumer Price Index for all urban consumers nationally.  Input levels for the variable costs were then scaled relative to the number of fishing hooks deployed to enable efficient optimization during model calibration and simulations.  Prices of inputs were adjusted using the inverse scaling ratio to preserve the observed expenditure for each input.  We were able to match vessels across data sources using vessel name, permit number, and commercial license. 

In 2012, there were 129 vessels operating in Hawaii's longline fishery.  Of the 129 vessels operating, 114 were represented in the cost and earnings survey (\cite{Pan2015a}).  We imputed input cost for missing vessels using random regression imputation considering gear usage, vessel catch profile, and time spent on each target as regression variables.  Variable costs were then grouped into six categories: fuel, captain pay, crew pay, bait, other, and gear.  We grouped fuel and oil costs under fuel, fixed captain pay and shares paid to the captain under captain pay, combined crew fixed pay and crew shares paid under crew pay, total bait costs under bait, and gear replacement cost under gear.  Table \ref{tab1} shows the degree of fleet heterogeneity based on these inputs.  According to the survey data, total variable costs exceeded total gross revenue for six vessels.  Rather than dropping these vessels because they violated the profit maximizing assumption, we scaled their input costs such that annual profits were 0.

We then disaggregated individual vessel expenditure, catch, and revenue by three policy relevant targets: bigeye EPO, bigeye WCPO, and swordfish.  The EPO and WCPO management regions are separated at 150 W longitude.  Bigeye and swordfish fishing sets differ by depth, with swordfish lines set shallower than deep set bigeye lines.  We used set-type and location from 2012 logbook data to calculate the proportion of total trip time spent each trip on each target.   Trip target time was then aggregated by vessel over the entire year indicating how much time each vessel spent on each target for 2012.  Using the dealer data from 2005-2013, we matched vessel trips to observed landings to calculate annual catch and revenue by vessel and target.  Observations in the dealer data recorded daily sales.  Fish sales were either recorded by individual fish or groups of fish sold together.  Daily vessel revenue was calculated by multiplying pounds sold per fish, or group of fish by recorded ex vessel price per pound.  The data were then aggregated on vessel and year to calculate the annual pounds of swordfish and bigeye caught, and the total value of vessel catch.  These data were then used to calculate fleet-wide average price of swordfish and bigeye, vessel-specific price premium for swordfish and bigeye, and price of non-target catch representing its added value.  Input expenditures for each vessel were disaggregated by target according to the proportion of time spent on each target in 2012.  Table \ref{tab2} summarizes the total active fleet size, and model sample size for each target over the years 2005-2013.

\begin{sidewaystable}[]
\centering
\caption{Time series data summary of total active and modeled vessels from the 2005 to 2013 dealer data.  Because some vessels fish more than one target, total vessels modeled can be less than the sum of each target.}
\label{tab2}
\resizebox{\textwidth}{!}{%
\begin{tabular}{llllll}
Year & Total Vessels Operating & Total Vessels Modeled & Vessels modeled (WCPO) & Vessels modeled (EPO) & Vessels modeled (SF) \\
2005 & 125                      & 105                   & 103                    & 41                    & 11                   \\
2006 & 127                      & 112                   & 111                    & 11                    & 10                   \\
2007 & 129                      & 116                   & 115                    & 53                    & 13                   \\
2008 & 129                      & 118                   & 115                    & 79                    & 11                   \\
2009 & 127                      & 120                   & 118                    & 73                    & 15                   \\
2010 & 124                      & 119                   & 115                    & 86                    & 15                   \\
2011 & 129                      & 124                   & 122                    & 83                    & 16                   \\
2012 & 129                      & 128                   & 127                    & 94                    & 17                   \\
2013 & 135                      & 126                   & 124                    & 83                    & 10                  
\end{tabular}}
\end{sidewaystable}

\section{Model specification}
The PMP framework consists of an objective function defining profit maximization and resource and policy constraints that restrict input allocation decisions.  To allow for non-linearity in production and limited substitution between inputs we chose to use a generalized constant elasticity of substitution (CES) production function, and for simplicity a linear expenditure function.  When paired with a CES production function, the linear expenditure function allows for smooth responses to changes in policy and resource constraints without adding more parameters to calibrate.  We define subscript  to index the set of 128 vessels in our sample,  indexes targets EPO, WCPO, and SF, and  indexes inputs for fuel, captain pay, crew pay, bait, other and gear.  Given a CES specification the production function for vessel  targeting  is given below.

\begin{equation}
y_{i,r}=\alpha_{i,r}(\sum_j \beta_{i,j,r} x_{i,j,r}^\rho)^{\delta/\rho}.
\end{equation}
 We define the scale parameter for vessel technology as $\alpha_{i,r}$, input share as $\beta_{i,j,r}$, elasticity of substitution as $\rho$, and the returns to scale coefficient as $\delta$.  By relating effort to catch, the scale parameter is analogous to a vessel-specific catchability parameter in traditional fishery production models.  The returns to scale coefficient is defined using a myopic definition (\cite{garnache2015calibration}) relating returns to scale to supply elasticity ($\eta$)  

\begin{equation}
\delta_{myo}=\frac{\eta}{1+\eta}.
\end{equation}
Because there have been no direct estimates of supply elasticity of catch in Hawaii's longline fleet, we assume $\eta=0.5$, which lies in the range of published supply elasticity estimates for the Gulf of Mexico fishery (\cite{zhang2011estimation}). To simplify notation, we use a transformed elasticity of substitution defined as

\begin{equation}
\rho=\frac{\sigma -1}{\sigma},
\end{equation}
where the untransformed elasticity of substitution ($\sigma$) is assumed to be 0.17 for all inputs. At present, we are unable to estimate an elasticity of substitution from the data available, and the value of 0.17 allows for limited substitution between inputs, which we borrow from the agriculture literature and feel is reasonable in a fishery setting (\cite{howitt2012calibrating}).  Model sensitivity analyses for these assumptions are provided in Figure \ref{figA1} and indicate our results are robust to changes in assumed parameter values.

Although our production function only models targeted catch, fisher's revenue will depend on their ability to land quality fish, and on the value of non-target but commercially valuable bycatch.  To fully capture these components of revenue we model the price of swordfish and bigeye separately for each vessel.  The fleet-wide average prices for swordfish and bigeye are given by $p_{i,sf}$, and $p_{i,be}$, vessel-specific price premiums for swordfish and bigeye accounting for variation in quality are given by $p_{i,sfpr}$, and $p_{i,bepr}$, and the additional values from non-targeted bycatch are given by $p_{i,nsf}$, and $p_{i,nbe}$.  By adding these three components together, we specify a vessel-specific price for bigeye (BE), and swordfish (SF).

\begin{equation}
p_{i,SF}=p_{i,sf}+p_{i,sfpr}+p_{i,nsf}
\end{equation}
\begin{equation}
p_{I,EPO}=p_{i,WCPO}=p_{i,be}+p_{i,bepr}+p_{i,nbe}
\end{equation}
This specification allows us to exactly reproduce observed vessel revenue, while only modeling the production of the policy relevant targets.  Implicit in this price specification we assume the price of bigeye from the EPO is the same as bigeye from the WCPO, which we feel is reasonable given they belong to the same species and are both caught throughout the year. 

For simplicity, we specify a linear expenditure function.  The input cost data only provides total annual costs per input, therefore we assume input prices ($c_{i,j,r}$) are 1, which implies input levels ($x_{i,j,r}$) are in dollar units.  The choice set $x_{i,j,r}$ is the vector of individual vessel input levels for each target.  Profit maximization is constrained by three policies.  We model annual catch limits for bigeye tuna in the EPO ($ACL_{EPO}$) and WCPO ($ACL_{WCPO}$), and a total annual catch limit for swordfish ($ACL_{SF}$).  Vessel heterogeneity implies that the unobserved value of catch for each constraint will vary by vessel.  We therefore define the unobserved value of catch as $\mu_{i,r}$ over vessels and targets.  The maximization problem is given below.

\begin{equation}
\begin{array}{rrclcl}
\displaystyle \max_{x_{i,j,r}} & \multicolumn{3}{l}{\sum_i \sum_r \left[ (p_{i,r}+ \mu_{i,r})y_{i,r} - \sum_j c_{i,j,r}x_{i,j,r} \right]} \\
\textrm{s.t.} & \sum_i y_{i,EPO} & \leq & ACL_{EPO} \\
& \sum_i y_{i,WCPO} & \leq & ACL_{WCPO} \\
& \sum_i y_{i,SF} & \leq & ACL_{SF} \\
\end{array}
\end{equation}

\section{Model calibration}
We adapted the calibration procedure developed by \cite{garnache2015calibration}.  Their calibration procedure is the most recent methodological advance in the PMP literature, comprehensively addressing the criticism by \cite{heckelei2003estimation} regarding the calibration of shadow values.  Rather than estimated using an LP or ad hoc measures as was done previously, all unknown parameters and the shadow values are calibrated simultaneously using the same structural forms as used in model simulations, in this case a CES production function with a linear expenditure function.  \cite{garnache2015calibration} calibrated a PMP model for agriculture. In agriculture, the constrained input is typically land, however, in fisheries, production inputs can be purchased at any desired level on a common market and the constrained resource is catch.  We adapted the calibration procedure to account for this difference. For each target we specified a shadow value ($\lambda_r$).  We then calibrated the model by minimizing the sum of squared error between observed expenditures and model expenditures resulting from the choice variable $\lambda_r$ as specified below. 

\begin{equation}
\min_\lambda \sum_i \sum_r \left[ (p_{i,r}+\lambda)\bar{q}_{i,r}\delta - \sum_j c_{i,j,r} \right]^2
\end{equation}

The objective function is subject to four sets of constraints that determine the calibration of unknown parameters.   The first set of constraints requires production parameters reproduce observed output ($\bar{q}_{i,r}$) for each vessel and target.

\begin{equation}
\bar{q}_{i,r}=\alpha \left( \sum_j \beta_{i,j,r} x_{i,j,r}^\rho \right)^{\delta/\rho} , \forall i, r
\end{equation}
The second set of constraints requires the first order conditions of profit maximization hold. The first order condition will be specified for each input, vessel, and target as below.

\begin{equation}
\begin{split}
p_{i,r}\alpha_{i,r}\delta \left( \sum_j \beta_{i,j,r} x_{i,j,r}^\rho \right)^{(\delta/\rho) - 1} \beta_{i,j,r} x_{i,j,r}^{\rho-1} \\
= c_{i,j,r} - (\lambda+\mu_{i,r})\alpha \delta \left( \sum_j \beta_{i,j,r} x_{i,j,r}^\rho \right)^{(\delta/\rho)-1} \beta_{i,j,r} x_{i,j,r}^{\rho-1}, \forall i,j,r
\end{split}
\end{equation}
The third set of constraints allows us to recover the vessel and target-specific unobserved value of catch ($\mu_{i,r}$).

\begin{equation}
\begin{split}
p_{i,r}\sum_j \left[ \alpha \delta \left( \sum_j \beta_{i,j,r} x_{i,j,r}^\rho \right)^{(\delta/\rho)-1} \beta_{i,j,r} x_{i,j,r}^{\rho-1} \right] \\
= \sum_j c_{i,j,r} - (\lambda+\mu_{i,r})\alpha \delta \sum_j \left[ \left( \sum_j \beta_{i,j,r} x_{i,j,r}^\rho \right)^{(\delta/\rho)-1} \beta_{i,j,r} x_{i,j,r}^{\rho-1} \right], \forall i,j.
\end{split}
\end{equation}
Finally, our calibration procedure requires that for each vessel-target combination the sum of the input share parameters is one.

\begin{equation}
\sum_j \beta_{i,j,r} = 1, \forall i,r.
\end{equation}

\section{Calibration results}
The PMP model calibration procedure is designed to calibrate unknown parameters and constraint shadow values such that profit maximizing vessels, subject to the base year resource constraints, will optimally allocate the observed base year levels of input, generating the observed outputs and revenues, and the observed expenditures.  To evaluate whether the calibration was successful, we examine the range of calibrated parameter values and the differences between the observed and the modeled input levels using the base year catch constraints in 2012.

In Table \ref{tab3}, we present the range of calibrated model parameters. The largest magnitude of variation is found in unobserved shadow prices of catch and the scale parameters.  These parameters carry the most weight for modeling the heterogeneous responses of the fleet.  The share parameters also show significant variation indicating the model captured a large amount of vessel heterogeneity in input expenditures.   Across targets, the share parameter for fuel are consistently larger than the other inputs, which is expected given fuel is the largest single input cost.  To verify the calibration procedure, we examine the differences between observed and modeled input levels for each input and each vessel's output using the base year constraints.  The largest difference in input is 2.02x10-14\% and the largest difference in output is 9.53x10-6\%. Such small differences indicate that we achieve an accurate calibration of all unknown parameters, and that our model can very closely replicate the observed base year economic behavior of each vessel.

\begin{table}[]
\centering
\caption{Summary of calibrated parameters for vessels modeled vessel and target-specific PMP model.  The mean and standard deviation for each target-specific parameter are given.}
\resizebox{\textwidth}{!}{%
\label{tab3}
\begin{tabular}{lllll}
Description                     & Symbol & WCPO              & EPO             & SF                \\
Scale parameter                 &$\alpha$& 1,104.02 (418.80) & 382.38 (305.60) & 1,629.79 (496.58) \\
Shadow value                    &$\lambda$& -7.70 (NA)        & -7.57 (NA)      & -4.45 (NA)        \\
Unobserved price of catch       &$\mu$& 17.42 (4.65)      & 24.00 (16.33)   & 10.41 (2.32)      \\
Share parameter for fuel        &$\beta_{fuel}$& 0.42 (0.10)       & 0.57 (0.24)     & 0.89 (0.16)       \\
Share parameter for captain pay &$\beta_{cap}$& 0.19 (0.08)       & 0.14 (0.10)     & 0.04 (0.07)       \\
Share parameter for crew pay    &$\beta_{crew}$& 0.12 (0.09)       & 0.08 (0.08)     & 0.02 (0.02)       \\
Share parameter for bait        &$\beta_{bait}$& 0.13 (0.03)       & 0.10 (0.06)     & 0.02 (0.04)       \\
Share parameter for other       &$\beta_{other}$& 0.08 (0.03)       & 0.07 (0.04)     & 0.02 (0.03)       \\
Share parameter for gear        &$\beta_{gear}$& 0.05 (0.02)       & 0.04 (0.02)     & 0.02 (0.02)      
\end{tabular}}
\end{table}

To further verify the calibration procedure, we compare the shadow values to the observed average price per pound of fish.  The shadow value on each resource constraint can be interpreted as the value of relaxing the resource constraint by one pound of either bigeye or swordfish.  Taken in absolute value terms, the calibrated shadow values of -7.70, -7.57, and -4.45, representing bigeye catch in the WCPO, EPO, and swordfish catch respectively, appear to be accurately calibrated.  When compared to the average observed price per pound of bigeye, and swordfish (\$7.99, \$4.30 respectively), our calibrated shadow values are within a few cents of the average observed fish prices.  Although average prices and shadow values do not share the same interpretation, comparing the two does provide a useful validation of the overall calibration procedure.

\section{Prediction accuracy}
\begin{sidewaysfigure}
  \includegraphics[width=\linewidth]{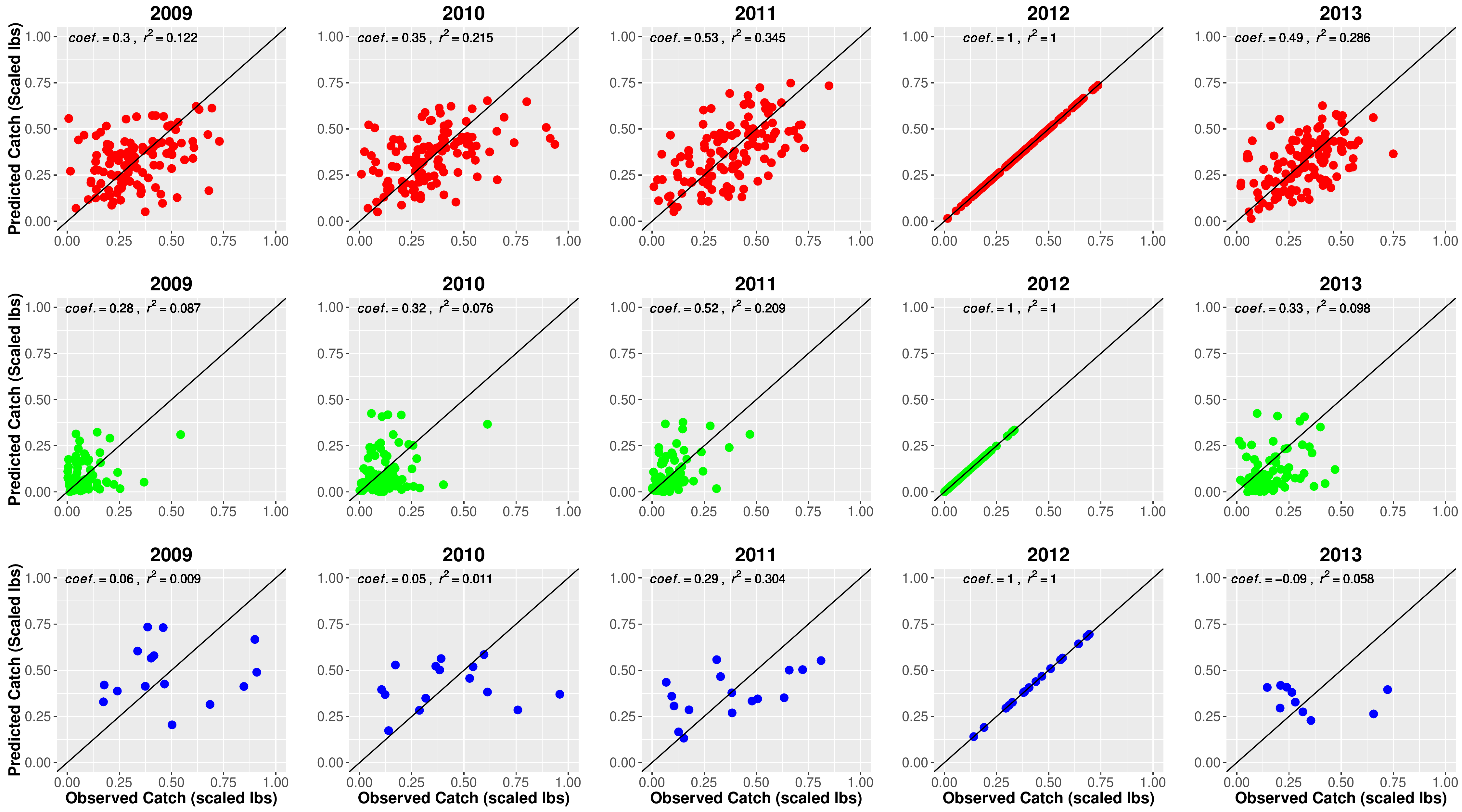}
  \caption{Evaluation of model predictions of individual vessel catch for bigeye in the WCPO (red), bigeye in the EPO (green), and swordfish (blue) from 2009-2013.  The solid line indicates the 45-degree line.  The correlation coefficient and R-Squared from the linear model are given in the top-left corner of each plot.  Axes are scaled so the maximum catch is 1 to prevent disclosure of confidential data.}
  \label{fig1}
\end{sidewaysfigure}

We evaluate model predictions in two ways.  First, we compare predicted and observed catch from 2009 to 2013.  Of the 128 vessels modeled, 126 were operating in 2013; however, going back to 2009, as few as 119 of the original 128 were previously operating (Table \ref{tab2}).  For each year, we simulate the model by setting the fleet-wide catch constraint less than or equal to the total observed catch of the vessels remaining from our 2012 sample.  This implies that our simulated fleet size decreases as vessels operating in 2012 are no longer observed in more distant years.  To account for changes in input costs over time, we adjust the cost of fuel using U.S. number 2 diesel retail price\footnote{https://www.eia.gov/dnav/pet/hist/LeafHandler.ashx?n=PET\&s=EMD\_EPD2D\_PTE\_NUS\_DPG\&f=A} and the costs for captain pay and crew pay using annual salary data from Bureau of Labor Statistics occupational profiles for farming, fishing, and forestry occupations.\footnote{http://www.bls.gov/oes/tables.htm} Regressing the predicted revenue on observed revenue for the years 2009-2013, we examine the correlation coefficient and the amount of variation explained by our model (Figure \ref{fig1}).  We find the model performs best predicting bigeye catch in the WCPO, modestly for bigeye catch in the EPO, and poorly for swordfish catch.  The best out-of-sample model predictions are made for the 2011 bigeye catch in the WCPO (R-squared=0.35, correlation coefficient=0.53).  For all targets, model predictions become less accurate moving further in time away from the calibrated base year.  This is expected as biological stock level, individual fishing location decisions, and environmental conditions could vary substantially over this time, while our model assumes conditions remain constant.  In the short-term the model makes reliable predictions of individual vessel catch for the largest target in the fishery, bigeye in the WCPO.  

Second, we evaluate the model input level predictions for each target comparing the observed input levels from the 2005 cost and earnings data to the predicted input levels simulated using our PMP model. Results are shown in Table 4.  In order to compare the values, we match vessels that appear in both sets, reducing our sample to 71, 25, and 1 for the WCPO, EPO, and SF targets respectively.  Results from a paired Wilcoxon test comparing the observed and predicted input expenditures show the model significantly under-predicts all inputs except gear and bait for the WCPO target.  The model tends to over-predict input costs for the EPO target, and it over-predicts all inputs except fuel for the one matched vessel targeting SF.  By comparing observed expenditures in 2012 (Table \ref{tab1}) to 2005 (Table \ref{tab4}), the primary source of prediction error is the large differences in the observed expenditures between 2012 and 2005.  For instance, fundamental changes to the remuneration schemes over these years, including the wide-spread transition from crew shares paid to domestic crew to fixed pay for foreign crew, could account for the observed differences in crew pay and captain pay.  We also observed a reduction in fuel expenditures in 2005 in the WCPO and EPO, and increase in SF, which could reflect a change in fishing grounds requiring more or less travel time than in 2012.  Similar explanations could account for differences in other input expenditures predicted for each target.  Gear and bait expenses, which we expect to be most closely tied to catch, generate the closest predictions and are not sensitive to changes in remuneration scheme or fishing location.  Any changes to the fundamental cost structure of the fleet are expected to alter model parameter values and reduce the accuracy of forecasts.  This limitation is common to all model based forecasts.

\begin{table}[]
\centering
\caption{Data summary of annual input costs in dollars for WCPO, EPO, and SF targets from the Cost and Earnings Survey in 2012.}
\label{tab1}
\begin{tabular}{llll}
Inputs      & Mean WCPO (SD)     & Mean EPO (SD)     & Mean SF (SD)      \\
Fuel        & \$154,045 (62,542) & \$27,134 (31,917) & \$16,318 (44,331) \\
Captain Pay & \$75,700 (47,061)  & \$13,623 (18,167) & \$6,962 (19,937)  \\
Crew Pay    & \$47,255 (46,103)  & \$7,245 (12,246)  & \$1,978 (6,192)   \\
Bait        & \$48,722 (17,761)  & \$7,928 (8,635)   & \$4,013 (10,787)  \\
Other       & \$31,477 (12,796)  & \$5,029 (5,652)   & \$3,195 (8,844)   \\
Hooks       & \$19,346 (8,583)   & \$3,160 (3,479)   & \$2,062 (5,618)  
\end{tabular}
\end{table}

\begin{sidewaystable}[]
\centering
\caption{Mean observed and the median difference between observed and predicted input expenditures.  Observed data came from the 2005 Cost and Earnings Survey.  All values are adjusted to 2012 dollars.  The median difference and p-values are from a two-sample paired Wilcoxon test.}
\label{tab4}
\resizebox{\linewidth}{!}{%
\begin{tabular}{lllllll}
            & WCPO                    &                                       & EPO           &                                         & SF            &                                       \\
Inputs      & Mean Observed (dollars) & Median Predicted Difference (P-value) & Mean Observed & Median Predicted   Difference (P-value) & Mean Observed & Median Predicted Difference (P-value) \\
            &                         &                                       & (dollars)     &                                         & (dollars)     &                                       \\
Fuel        & 106,532                 & -25,324                               & 10,972        & 9,388                                   & 22,349        & -4,563                                \\
            &                         & (\textless0.001)                      &               & -0.059                                  &               & (NA)                                  \\
Captain Pay & 84,114                  & -19,017                               & 7,404         & 5,789                                   & 11,937        & 4,229                                 \\
            &                         & -0.038                                &               & -0.101                                  &               & (NA)                                  \\
Crew Pay    & 56,204                  & -27,395                               & 5,150         & 3,523                                   & 9,744         & 6,471                                 \\
            &                         & (\textless0.001)                      &               & -0.022                                  &               & (NA)                                  \\
Bait        & 39,544                  & 45                                    & 3,627         & 5,141                                   & 8,312         & 7,561                                 \\
            &                         & -0.984                                &               & -0.007                                  &               & (NA)                                  \\
Other       & 32,259                  & -5,599                                & 2,635         & 2,920                                   & 4,785         & 10,807                                \\
            &                         & -0.011                                &               & -0.011                                  &               & (NA)                                  \\
Gear        & 17,426                  & -1,136                                & 1,430         & 2,153                                   & 4,006         & 3,866                                 \\
            &                         & -0.389                                &               & (\textless0.001)                        &               & (NA)                                  \\
Sample      & 71                      &                                       & 25            &                                         & 1             &                                      
\end{tabular}}
\end{sidewaystable}

\begin{figure}
  \includegraphics[width=\textwidth]{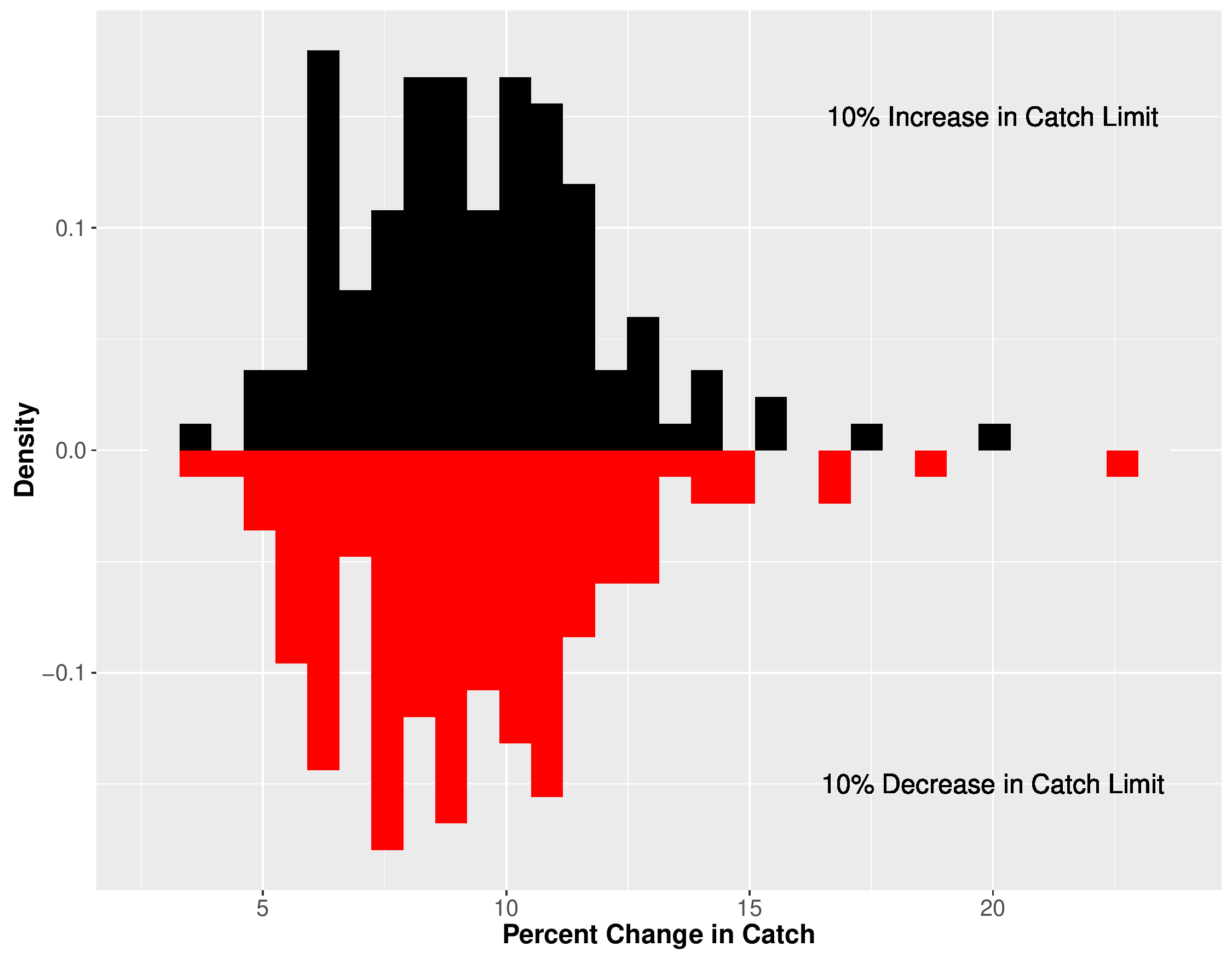}
  \caption{Distribution of responses for individual vessels measured by the percent change from 2012 catch levels.  Results from 10\% increase in annual catch constraint from 2012 are given filled black and represent increases in catch. Results from 10\% decrease in annual catch constraint from 2012 are filled red and represent decreases in catch.}
  \label{fig2}
\end{figure}

\section{Policy simulations}
To demonstrate the usefulness of a vessel-specific PMP model for Hawaii's longline fishery, we examine vessel responses and impacts on individual vessel catch to changes in the annual catch limit policy.  We simulate two policy changes.  The first is a policy that increases the annual catch limit of bigeye in the WCPO by 10\% from the 2012 base year.  The second is a policy that decreases the same catch limit by 10\% from the 2012 base year.  A 10\% change in the catch limit policy is roughly in line with the agreed upon changes for bigeye in the WCPO in the next few years which will see catch limit decrease 11\% from 3,763 metric tons in 2014, to 3,345 metric tons in 2017.

The vessel-specific nature of our PMP model allows us to evaluate the distributional effects of such policy changes.  We expect that individual vessels will respond to varying degrees, depending on factors such as technological efficiency and profitability, which makes them more or less sensitive to policy changes.  In Figure \ref{fig2}, we present the distribution of catch responses given an increase and decrease in bigeye catch limits in the WCPO.  The range of responses is large.  With a 10\% increase in catch limit, we see that vessels respond by increasing catch from less than 5\% to 20\%.  With a 10\% decrease in catch limit, the responses are symmetric to the 10\% increase policy.  Vessels reduce catch from less than 5\% to 25\%.  Given the range in policy responses, individual vessels will clearly be affected differently.  Some will be highly sensitive to policy changes; most will experience moderate impacts.  Understanding the distributional implications is clearly important for evaluating economic impacts of fishery policies in Hawaii's longline fishery.

\section{Conclusion}
In this paper, we have shown that the vessel and target specific PMP model of Hawaii's longline fishery reliably predicts short-term effect of policies on bigeye catch in the WCPO and EPO.  Model predictions are more accurate when simulating vessel responses close to the base year, but lend some insight even at further distances.  By calibrating at the vessel-specific level, we are able to identify the range of economic responses to policy changes, capturing the heterogeneous nature of Hawaii's longline fleet.  This more realistically models vessel responses, as well as provides an evaluation of the distributional effects of policy changes on catch, which is important for evaluating the stability of new policies.  For fishery managers, the PMP model of Hawaii's longline fishery provides a valuable tool for evaluating the economic impacts of current and potential fishery policies. 

The PMP framework also provides a rich structural model with which we can study fisheries in general.  Later work will address parameter instability resulting from fundamental changes to underlying economic relationships or environmental and biological conditions, and estimate target switching decisions made by fishers.  We will also consider the effects of overlapping policy constraints such as turtle interaction caps, and explore the individual vessel characteristics that make certain vessels more sensitive to policy changes than others.


\printbibliography

\section{Appendix}

\begin{figure}
  \includegraphics[width=\textwidth]{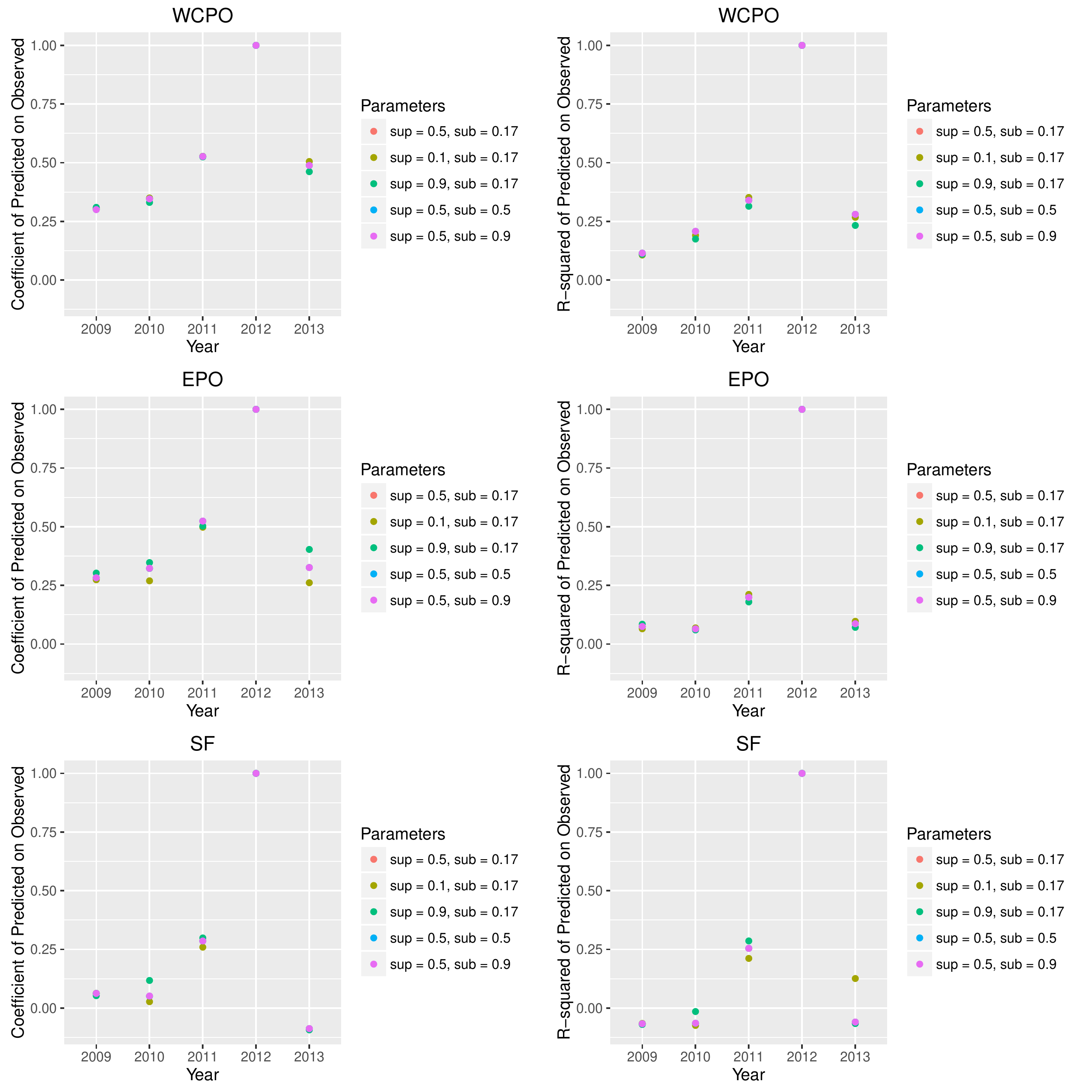}
  \caption{Sensitivity analysis measuring the effect from changing assumed supply elasticity and substitution elasticity values on model prediction results from 2009-2013.}
  \label{figA1}
\end{figure}

\end{document}